\documentclass[twocolumn,secnumarabic,amssymb, nobibnotes, aps, pre]{revtex4-1}

\usepackage{amsmath}
\usepackage{amssymb}
\usepackage{amsthm}
\usepackage{graphicx}
\usepackage{listings}
\usepackage{setspace}
\usepackage{psfrag}
\usepackage{float}

\newcommand{\pa}{\partial}

\begin{document}

\title{Localised Oscillatory States in Magnetoconvection}%

\author{Matthew~C.~Buckley}%
\author{Paul~J.~Bushby}%
\affiliation{School of Mathematics and Statistics, Newcastle University, Newcastle Upon Tyne, UK}%
\date{October 29, 2012}%
\begin{abstract}
Localised states are found in many pattern forming systems. The aim of this paper is to investigate the occurrence of oscillatory localised states in two-dimensional Boussinesq magnetoconvection. Initially considering an idealised model, in which the vertical structure of the system has been simplified by a projection onto a small number of Fourier modes, we find that these states are restricted to the low $\zeta$ regime (where $\zeta$ represents the ratio of the magnetic to thermal diffusivities). These states always exhibit bistability with another non-trivial solution branch, in other words they show no evidence of subcritical behaviour. This is due to the weak flux expulsion that is exhibited by these time-dependent solutions. Using the results of this parameter survey, we locate corresponding states in a fully-resolved two-dimensional system, although the mode of oscillation is more complex in this case. This is the first time that a localised oscillatory state, of this kind, has been found in a fully-resolved magnetoconvection simulation.  
\end{abstract}

\maketitle
%\tableofcontents

\section{\label{sec:Introduction}Introduction}
Localised states are found in a wide range of physical systems. Examples include the localised buckling patterns that are observed in the bending of beams \cite{MWLG2012}, cavity solitons in a semiconductor-based optical amplifier \cite{SB2008}, isolated current filaments in a gas discharge system \cite{YAYL1997} and soliton-like structures on the surface of a ferrofluid \cite{RRIB2005}. Oscillatory localised states (`oscillons') have been observed experimentally in vertically-vibrated media \cite{PBU1996,LHARF1999}. With appropriate choices for the nonlinear terms, the Swift-Hohenberg equation exhibits a range of localised stable solutions in both one \cite{JBEK2006,JBEK2007,GKPASC2009,JBJD2012} and two spatial dimensions \cite{HSHB1997,CCHR1999}, whilst equations of Ginzburg-Landau type can also produce localised states \cite{JBAYEK2008,JDSL2010}. Spatially-localised behaviour has also been found in simulations of two-dimensional thermosolutal convection \cite{KGAM1997,ABEK2008a,ABEK2008b,CBABEK2011} as well as in two-dimensional simulations of convection in a binary fluid \cite{OBEK2005,BKAM2006,MBAK2011}. 

The presence of a strong vertical magnetic field tends to inhibit convective motions in an electrically-conducting fluid \cite{MPNW1982}. This phenomenon can be observed on the surface of the Sun, where the dark central part of a sunspot (which is usually referred to as the umbra) corresponds to a region of the solar surface in which the convective heat transport is impaired by an intense concentration of vertical magnetic flux. However, observations suggest that convection is not completely suppressed in these regions. In particular, there are some small bright features within sunspot umbrae, usually known as umbral dots \cite{RK2007,JTNW2008}, which tend to be non-uniformly distributed across the umbral region \cite{MSAH2005}. It is probable that umbral dots indicate the presence of localised magnetoconvective motions. These observations have motivated theorists to search for localised states in idealised models of magnetoconvection. 

It is now well-established that it is possible to find steady localised states (often referred to as ``convectons") in two-dimensional Boussinesq magnetoconvection \cite{SB1999,JD2007}. However, the bifurcation structure of this magnetoconvection system differs from that exhibited by simpler systems, such as the illustrative Swift-Hohenberg models, where localised states exist in a region of bistability between the conducting state and the multiple roll, spatially-periodic convective pattern \cite{JBEK2006,JBEK2007}. Furthermore, in the Swift-Hohenberg case, solution branches corresponding to localised states form a vertical ``snaking" pattern of interlaced curves of gradually increasing numbers of oscillations \cite{GKSC2006,SCGK2009}. This snaking pattern is mostly confined to a narrow region of parameter space, usually referred to as the ``pinning" region. In a finite domain, this pattern of snaking solution branches eventually terminates on the periodic patterned solution branch. In the case of magnetoconvection models, the presence of a conserved quantity (namely the vertical magnetic flux) has an important effect upon the bifurcation structure \cite{PMSC2000}. In a highly conducting fluid, convective eddies tend to expel magnetic flux \cite{NW1966}, which must therefore accumulate elsewhere within the domain. This can lead to the complete suppression of convective motions in the magnetically-dominated parts of the layer, leading to a flux separated state \cite{LT1998} in which most of the convective motions are restricted to field-free regions. This flux expulsion process enables convectons to exist in a subcritical regime in which the conducting state is the only non-localised stable solution. Numerical continuation has been used to show that the solution branches corresponding to localised states in magnetoconvection, which tend to bifurcate subcritically from the periodic states, exhibit a slanted snaking pattern that is not confined to a narrow pinning region \cite{JD2008,DLJ2011,DLJ2012}. Although it has been suggested that standard snaking should be recovered in the limit as the box size tends to infinity \cite{DLJ2011,DLJ2012}, the existence of slanted snaking in finite domains illustrates the fact that the bifurcation structure of magnetoconvection differs in a fundamental way from that of many comparable pattern-forming systems. This is therefore an interesting system to study in its own right.

One of the notable features of Boussinesq magnetoconvection is that it is possible to find regions of parameter space in which the trivial conducting state is unstable to a spatially-periodic oscillatory mode \cite{MPNW1982}. This raises the question of whether it might be possible to find localised oscillatory states \cite{EK2008}. Oscillatory convectons have been found in both two \cite{SB1999b} and three \cite{SBNW2002} spatial dimensions, but only in idealised models in which the vertical structure of the system has been simplified by a projection onto a small number of Fourier modes. Little is known about the parametric dependencies of these states in two-dimensions and, so far, the existence of such states has been confirmed only in a very small region of parameter space (using the parameters defined below, $R=20000$, $\zeta=0.1$, $\sigma=1$, $\lambda=6$ and $Q=17000-22000$ \cite{SB1999b}). Whether or not fully-resolved magnetoconvection can produce corresponding oscillatory states is still an open question. As described above, only steady localised states have so far been found in fully-resolved two-dimensional Boussinesq magnetoconvection. This also seems to be the case in fully-resolved three-dimensional compressible magnetoconvection \cite{SHPB2011}, although, again, only a limited exploration of parameter space has been carried out to date. So, on the basis of previous studies, we cannot discount the possibility that oscillatory localised states exist only in simplified models of magnetoconvection. Having said that, given the complexity of the bifurcation structure for the steady modes it is just as plausible that these states do exist but have yet to be observed in numerical simulations.

The primary aim of this paper is to demonstrate the existence of oscillatory localised states in fully-resolved two-dimensional magnetoconvection. We focus initially upon a simplified model with a truncated vertical structure, with the aim of determining the properties of oscillatory localised states in this system. We also carry out a brief parametric survey in order to verify that these oscillatory localised states are not restricted to a narrow region of parameter space. So, in that sense, these solutions are a robust feature of this system. Having carried out this survey, we then move on to the more complex fully-resolved case. In the next section, we describe our magnetoconvection model. Numerical results are presented in Section 3. Finally, our conclusions are given in Section 4.

\section{\label{sec:Model}Model}
  \subsection{\label{sec:TFRM}The Fully Resolved Model}

Adopting the Boussinesq approximation \cite{ESGV1960,MPNW1982} we consider a layer of electrically conducting fluid that is heated from below, in the presence of an imposed uniform vertical magnetic field.  The fluid occupies a two-dimensional Cartesian domain of dimensions $0 \le z \le d$ and $0 \le x \le \lambda d$, with periodic boundary conditions being enforced in the horizontal direction.  The orientation of the Cartesian system of coordinates is chosen so that the $z$-axis points vertically upwards, which implies that the constant gravitational acceleration is given by ${\boldsymbol g} = -g \hat{{\boldsymbol z}}$.  The magnetic field is constrained to be vertical at the upper and lower boundaries, which are also assumed to be impermeable and stress-free, and are held at fixed uniform temperatures. Various properties of the fluid are assumed to be constant, including the thermal conductivity $\kappa$, the kinematic viscosity $\nu$, the magnetic permeability $\mu_0$, the thermal expansion coefficient $\bar{\alpha}$, and the (reference) density $\rho_0$. 
To non-dimensionalise this system, we follow the approach of previous authors \cite{MPNW1982} by scaling length and time by the layer depth, $d$, and the thermal relaxation time, $d^2/\kappa$, respectively. The velocity field $\boldsymbol{u}\left( {\boldsymbol x}, t \right)$ is therefore scaled by $\kappa/d$,  whilst the magnetic field $\boldsymbol{B}\left( {\boldsymbol x}, t \right)$ is scaled by the magnitude of the imposed field, $|\boldsymbol{B}_0|$. Finally, we normalise $T\left( {\boldsymbol x}, t \right)$, which is a measure of the temperature of the fluid relative to that of the upper boundary, by the temperature difference across the layer, $\Delta T$. The dimensionless governing equations can then be written in the following form
        \begin{align}
          \frac{1}{\sigma} \left( \frac{\pa {\boldsymbol u}}{\pa t} + \left( {\boldsymbol u} \cdot \nabla \right){\boldsymbol u} \right) &= -\nabla p + R T \hat{{\boldsymbol z}} \notag \\
&+ \zeta Q \left( \nabla \times {\boldsymbol B} \right) \times {\boldsymbol B} + \nabla^2 {\boldsymbol u}, \label{equ:NS} \\[1em]
          \frac{\pa T}{\pa t} + \left( {\boldsymbol u} \cdot \nabla \right) T  &= \nabla^{2} T, \label{equ:TE} \\[1em]
          \frac{\pa {\boldsymbol B}}{\pa t} - \nabla \times \left( {\boldsymbol u} \times {\boldsymbol B} \right) &= \zeta \nabla^{2} {\boldsymbol B}, \label{equ:IE}
        \end{align}
where $p\left(\boldsymbol{x}, t \right)$ is the (modified) pressure \cite{MPNW1982}, whilst the (incompressible) velocity field and the magnetic field satisfy the constraints 
        \begin{equation}\label{equ:conditions}
          \nabla \cdot {\boldsymbol u}=\nabla \cdot {\boldsymbol B}=0.
        \end{equation}
The four non-dimensional parameters in this system are defined in Table \ref{tbl:parameters}. The Chandrasekhar number, $Q$, is a measure of the dimensionless field strength. The Rayleigh number, $R$, measures the destabilising effect of the superadiabatic temperature gradient. The Prandtl number, $\sigma$, is the ratio of the viscous to thermal diffusivities, whilst $\zeta$ gives the ratio of the magnetic to thermal diffusivities.
\begin{table}[ht]
\centering                          
\begin{tabular}{c c c}            
\hline\hline \\[-2.0ex]                     
Symbol & Parameter Name & Definition\\
\hline  \\[-2.0ex] 
Q & Chandrasekhar number & $ |{\boldsymbol B}_{0}|^2 d^2/\mu_0 \rho_0 \nu \eta $\\
$R$ & Rayleigh number & $ g \bar{\alpha} \Delta T d^3/\kappa \nu $\\    
$\sigma$ & Prandtl number & $ \nu/\kappa $\\             
$\zeta$ & Diffusivity ratio & $ \eta/\kappa $\\
\hline                            
\end{tabular}
\caption{\label{tbl:parameters} The non-dimensional parameters in the governing equations for Boussinesq magnetoconvection.}
\end{table}
It is convenient to eliminate the pressure by taking the curl of Equation~(\ref{equ:NS}), which introduces the vorticity, ${\boldsymbol \omega}({\boldsymbol x},t) = \nabla \times {\boldsymbol u}$. Given that we are working in two spatial dimensions, the incompressibility condition for the velocity and the solenoidal constraint for the magnetic field (see Equation~(\ref{equ:conditions})) can be satisfied automatically by introducing a streamfunction, $\psi({\boldsymbol x},t)$:
    \begin{equation}
      {\boldsymbol u} = \nabla \times \left( \psi \left( x,z,t\right) \hat{{\boldsymbol y}} \right) = \left( -\pa_{z} \psi ,0, \pa_{x} \psi \right),
    \end{equation}
and a flux function, $A({\boldsymbol x},t)$:
    \begin{equation}
      {\boldsymbol B} = {\boldsymbol B_{0}} + \nabla \times \left( A\left(x,z,t \right) \hat{{\boldsymbol y}} \right) = \left( -\pa_{z} A ,0, 1+ \pa_{x} A \right),
    \end{equation}
where $ {\boldsymbol B_{0} =\hat{{\boldsymbol z}}}$ is the dimensionless, uniform vertical magnetic field \cite{SC1961,EK1981,MPNW1982}. We also define $\theta({\boldsymbol x},t)$, to be the temperature perturbation from the background conducting state, so that
    \begin{equation}
      T= 1-z+\theta\left(x,z,t\right).
    \end{equation}
This leads to the following set of governing equations
    \begin{align}
      \frac{\pa \omega}{\pa t} + \frac{\pa \left( \psi , \omega \right)}{\pa \left(x , z\right)} &= \sigma \nabla^{2} \omega - \sigma R \frac{\pa \theta }{\pa x} \notag \\
                &- \sigma \zeta Q \left(\frac{\pa \nabla^{2} A}{\pa z} + \frac{\pa \left(A,\nabla^{2} A\right)}{\pa \left(x, z\right)} \right), \label{equ:omega} \\[1em]
      \frac{\pa \theta}{\pa t} + \frac{\pa \left( \psi , \theta \right)}{\pa \left(x , z\right)} 
                &= \nabla^{2} \theta + \frac{\pa \psi }{\pa x}, \label{equ:theta} \\[1em]
      \frac{\pa A}{\pa t} + \frac{\pa \left( \psi , A \right)}{\pa \left(x , z\right)} 
                &= \zeta \nabla^{2} A + \frac{\pa \psi }{\pa z}, \label{equ:a}
    \end{align}
where $\omega(x,z,t)$ is the magnitude of the vorticity vector (which is parallel to the $y$-direction) and
    \begin{equation}\label{equ:poisson}
      \omega(x,z,t)=-\nabla^2 \psi(x,z,t).
    \end{equation}
Note we have used Jacobian notation in equations (\ref{equ:omega})-(\ref{equ:a}) where,
    \begin{equation}      
      \frac{\pa \left( f , g \right)}{\pa \left(x , z\right)} = \frac{\pa f}{\pa x}\frac{\pa g }{\pa z} - \frac{\pa f}{\pa z}\frac{\pa g }{\pa x},
    \end{equation}
with $f=f(x,z,t)$ and $g=g(x,z,t)$.  Having made this change of variables, the boundary conditions can be written in the following form:
    \begin{equation}\label{equ:bc3}
      \psi=\omega=\theta=\pa_z A=0 \hspace{0.1in} \text{at} \hspace{0.1in} z=0,1.
    \end{equation}
Equations (\ref{equ:omega})-(\ref{equ:a}) have a trivial equilibrium solution such that
    \begin{equation}\label{equ:trivial}
      \omega = \psi = \theta = A = 0,
    \end{equation}
which corresponds to a static fluid with a constant temperature gradient, permeated by a uniform, vertical magnetic field (i.e. the basic conducting state). 
 
  \subsection{\label{sec:TTM}The Truncated Model}
A number of previous studies of Boussinesq magnetoconvection have considered a further simplification to this model. The full horizontal structure is retained but, motivated by the fact that the convective pattern close to onset has a relatively simple vertical dependence, only a minimal (non-trivial) set of modes are retained in the vertical Fourier decomposition of each variable \cite{EK1981, SB1999, JD2007}. Following these authors, we satisfy the boundary conditions (\ref{equ:bc3}) by considering the following Fourier expansions,
    \begin{align}
      \psi(x,z,t) &= \psi_1 \left( x , t \right)\sin\left( \pi z \right), \\
      \omega(x,z,t) &= \omega_1 \left( x , t \right)\sin\left( \pi z \right), \\
      \theta(x,z,t) &= \theta_1 \left( x , t \right)\sin\left( \pi z \right) + \theta_2 \left( x , t \right)\sin\left( 2 \pi z \right), \\
      A(x,z,t) &= A_0 \left( x , t \right) + A_1 \left( x , t \right)\cos\left( \pi z \right).
    \end{align}   
It should be noted that both the $z$-dependent and $z$-independent components of the magnetic flux function are included in order to incorporate the flux expulsion effect \cite{NW1966, EK1981}. It is necessary to include $\theta_2$ in order to ensure that the hydrodynamic problem retains some form of nonlinearity. Projecting onto these modes we obtain the following set of five partial differential equations, which will become our one-dimensional truncated model \cite{JD2007}
    \begin{align}
      \pa_t \omega_1 &= \sigma \left(\omega_1''- \pi^2 \omega_1 \right) - \sigma R \theta_1' \notag \\
                     &- \sigma \zeta Q \pi \left[\left( 1 + A_0' \right)\left( \pi^2 A_1 - A_1'' \right) + A_0''' A_1 \right], \label{equ:w1cart} \\[1em]
      \pa_t \theta_1 &= \theta_1'' - \pi^2 \theta_1 + \psi_1' \left( 1 + \pi \theta_2 \right) + \frac{\pi}{2} \psi_1 \theta_2', \label{equ:theta1cart} \\[1em]
      \pa_t \theta_2 &= \theta_2'' - 4 \pi^2 \theta_2 + \frac{\pi}{2} \left( \psi_1 \theta_1' - \psi_1' \theta_1 \right), \label{equ:theta2cart} \\[1em]
      \pa_t A_0 &= \zeta A_0'' + \frac{\pi}{2} \left( \psi_1 A_1 \right)', \label{equ:A0cart} \\[1em]
      \pa_t A_1 &= \zeta \left( A_1'' - \pi^2 A_1 \right) + \pi \psi_1 \left( 1 + A_0' \right), \label{equ:A1cart}
    \end{align}
with the condition that 
    \begin{equation}\label{equ:psirel}
      \omega_{1}= \pi^2 \psi_{1} - \psi_{1}'',
    \end{equation}
(where primes denote $\pa_x$).

   \subsection{\label{ssec:CD}Code Details}

Due to the complexity of equations (\ref{equ:omega})-(\ref{equ:a}) and (\ref{equ:w1cart})-(\ref{equ:A1cart}) we must solve them numerically. This is accomplished by discretising the equations onto one and two-dimensional Cartesian meshes, with grid resolution $256$ and $256 \times 64$ grid points respectively. Both codes are pseudo-spectral, using Fast Fourier Transforms (FFT) from standard FFTW libraries to calculate all horizontal derivatives. In addition, the fully-resolved model uses fourth order finite differences to calculate all vertical derivatives. The discretised systems are evolved using a fourth order Runge-Kutta method for time-stepping with the timestep size constrained via the Courant-Friedrichs-Lewy (CFL) condition \cite{NR1986}. The vorticity and the stream function are related by Equation~(\ref{equ:poisson}) in the fully-resolved model, and by Equation~(\ref{equ:psirel}) in the truncated model. So, whenever the vorticity is evolved in time, these equations must be inverted in order to find the streamfunction. For the truncated model, this is accomplished by moving into Fourier space, where a second derivative corresponds to multiplication by $\left( i k \right) ^{2}$, where $k=2 \pi / \hat{\lambda}$ is the wavenumber and $\hat{\lambda}$ is the wavelength. Thus in Fourier space
    \begin{equation}
      \hat{\psi_{1}} = \frac{\hat{\omega_{1}}}{\pi^2+k^2},
    \end{equation}
where $\hat{\psi_{1}}$ and $\hat{\omega_{1}}$ correspond to the Fourier transforms of $\psi_{1}$ and $\omega_{1}$ respectively. For the fully-resolved case, we invert Equation~(\ref{equ:poisson}) using an LU-decomposition \citep{NR1986}, having already simplified the horizontal dependence of the problem by again moving into Fourier space.

  \section{\label{sec:Results}Numerical Results}

In this section we investigate the properties of oscillatory localised states in this magnetoconvection model, focusing initially upon the truncated model (in Section~\ref{ssec:OSTTM}) before moving on to the fully-resolved two-dimensional case (in Section~\ref{ssec:OSTFRM}). Before proceeding, we first define two important quantities. In what follows, we shall often refer to the Nusselt number:
	\begin{equation}\label{equ:nusselt}
	  N = \lambda^{-1} \int_{0}^{\lambda} \left( 1 -\pa \theta / \pa z \right) \,dx,
	\end{equation}
which measures the ratio of convective to conductive heat transfer \cite{MPNW1982}. We always measure this (often time-dependent) quantity at the base of the domain. Note that for $N=1$ we have a purely conductive state and thus no convective motions. For $N>1$, convection is occurring, with larger values corresponding to more vigorous convective motions.  For oscillatory states we shall see that the Nusselt number varies with time about a mean, given by $\bar{N}$. As described in the Introduction, convectons are associated with the phenomenon of flux expulsion. To quantify this effect, we define
\begin{equation}
      Q_{\text{eff}}=QB_{\text{z}}^2,
\end{equation}
where $B_{z}$ is the (uniform) field in the non-convective part of the layer. This is a measure of the effective field strength in the magnetically-dominated regions within which the localised state is embedded. When flux expulsion is very efficient, which is certainly the case for steady convectons, this definition is (approximately) equivalent to
\begin{equation}
 Q_{\text{eff}}= Q\lambda^2/\left( \lambda - \bar{\lambda} \right)^2,
    \end{equation}
where $\bar{\lambda}$ represents the cell width \cite{JD2007}.

    \subsection{\label{ssec:OSTTM}The Truncated Model}

      \subsubsection{\label{ssec:PDOOS}General Properties of the Oscillatory States}

      \begin{figure}
        \begin{center}
        \includegraphics[scale=0.14]{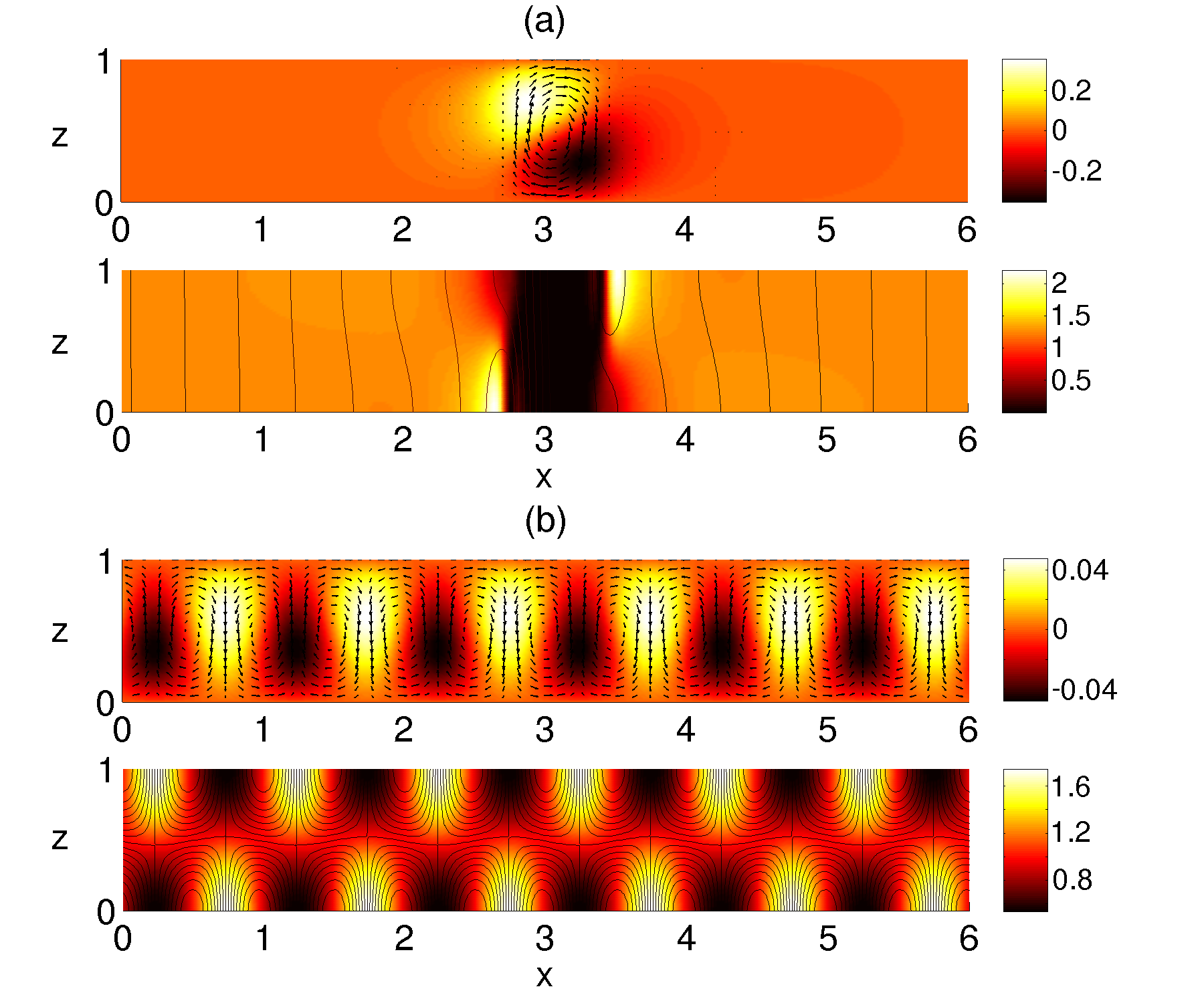}
        \caption{\label{fig:convoscill}(Color online) Two supercritical solutions obtained for the same parameter values, $R=20000$, $\zeta=0.1$, $\sigma=1$, $\lambda=6$ with $Q=22000$, for the truncated model, indicating the bistability of these states; (a) a steady convecton solution with $N=1.31$ and $\bar{\lambda}=0.996$ ($Q_{\text{eff}}=31631$) (b) 12-roll oscillatory state with $\bar{N}=1.08$. Each plot shows the temperature perturbation $\theta$ (top) and magnetic field strength, $|{\boldsymbol B}|^2$ (bottom).}
        \end{center}
      \end{figure}

      \begin{figure*}
        \begin{center}
          \includegraphics[scale=0.3]{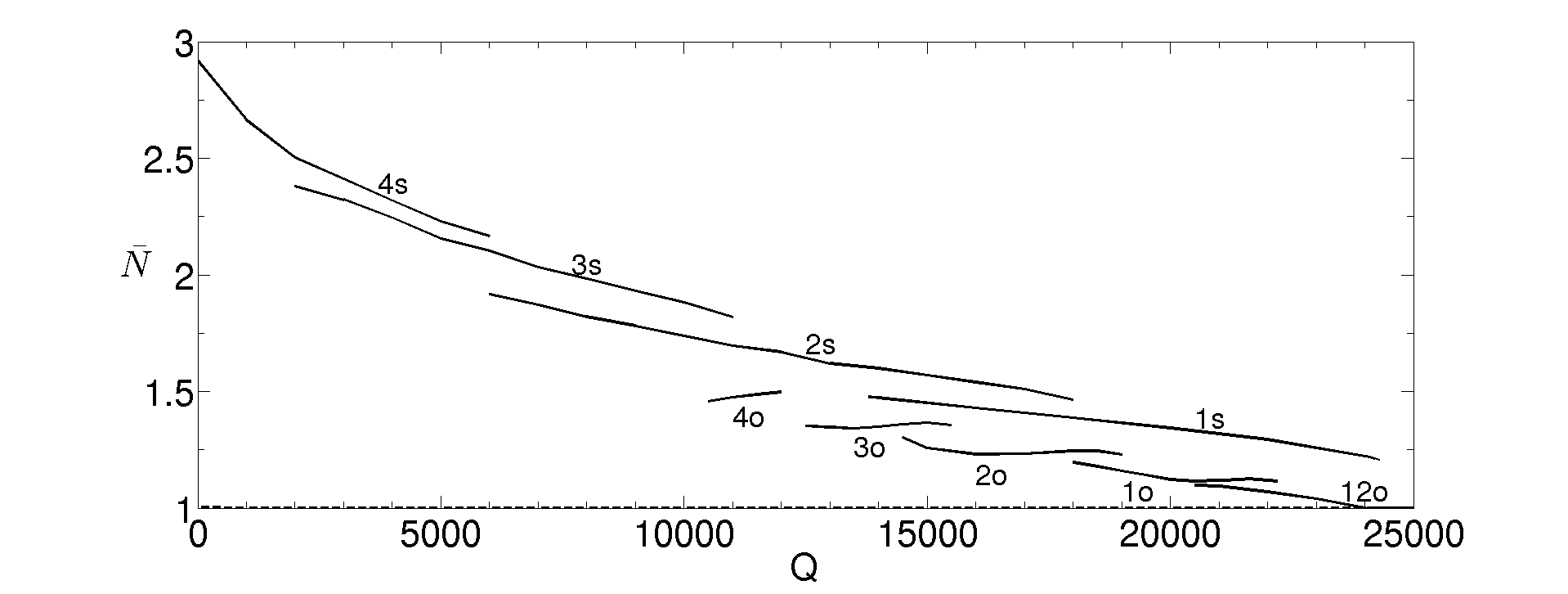}
          \caption{\label{fig:bifoscill} A bifurcation diagram indicating the stability range of a selection of stable steady (s) and oscillatory states (o) for the parameter values $R=20000$, $\zeta=0.1$, $\sigma=1$ and $\lambda=6$, where the dashed line (- -), at $N=1$, indicates the unstable region of the trivial state.  The numbers next to the branches refer to the number of convective rolls. }
        \end{center}
      \end{figure*}

Initially, we fix most of the parameters (setting $R=20000$, $\sigma=1$, $\zeta=0.1$ and $\lambda=6$) but vary the Chandrasekhar number, $Q$. It has been demonstrated that both steady and oscillatory localised states exist in the truncated model in this parameter regime \cite{SB1999b}, so this is a natural starting point for the present parameter survey. We fix $\sigma=1$ throughout, for numerical convenience. In a wide domain, the onset of convection in the absence of a magnetic field occurs at $R_c \approx 657.51$. So $R=20000$ corresponds to a highly supercritical value of the Rayleigh number for hydrodynamic convection. Of course convection will be completely suppressed if the Chandrasekhar number exceeds some critical value. As $Q$ is decreased from a very large value, the trivial conducting state eventually becomes unstable at $Q=Q_{\text{max}}^{\text{(o)}}=24047$ to an oscillatory convective mode (the initial bifurcation is always oscillatory provided that $\zeta<0.83$). The conducting state is also linearly-unstable to steady convective modes, although this occurs at much lower $Q$. Decreasing $Q$, the first steady bifurcation occurs at $Q=Q_{\text{max}}^{\text{(e)}}=1374 \ll Q_{\text{max}}^{\text{(o)}}$, so these bifurcations are not discussed further here. Henceforth, we define the subcritical regime to be the region of parameter space in which the trivial state is stable to convective perturbations (i.e. $Q>Q_{\text{max}}^{\text{(o)}}$), whilst the supercritical regime is defined by $Q<Q_{\text{max}}^{\text{(o)}}$. Previous studies suggest that steady localised states exist only in large aspect ratio domains, but $\lambda=6$ appears to be sufficient for this choice of parameters \cite{SB1999b}.  We will investigate the effects of varying $\lambda$ in the next subsection.
      
Before considering the properties of any oscillatory localised states that might be present, we first briefly review the properties of the steady convectons that exist in this system. These can be located by tracking solution branches \cite{SB1999}, starting from (for example) a steady multiple roll, flux-separated state, which can be obtained by evolving the system from the trivial (conducting) state at comparatively low $Q$. Using a flux-separated state as an initial condition, the value of $Q$ is gradually increased (typically using steps of $Q \approx 500-1000$), leading to progressively fewer convective rolls, until a single steady localised state is obtained. Note that it is necessary to take smaller step sizes (of $Q\approx 100$) close to the upper and lower stability boundaries of the convecton branches due to the fact that these states are more sensitive to small variations in $Q$ in these regions of parameter space. Each state was allowed to settle for approximately $1000$ thermal diffusion times, or until the Nusselt number exhibited no further time-dependence. One of these convecton states is illustrated in Fig.~\ref{fig:convoscill}(a). The upper plot in this image shows the temperature perturbation, and the lower plot shows the magnetic field strength, where lighter colours correspond to warmer fluid and stronger fields and darker colours correspond to cooler fluid and weaker fields. Steady convectons are very efficient at expelling magnetic flux from their interior, which causes the effective field strength of the outside region to increase. In the solution that is illustrated in Fig.~\ref{fig:convoscill}(a), $Q_{\text{eff}}=31631>Q_{\text{max}}^{\text{(o)}}$. Although the convecton branch does extend into the subcritical regime for this particular choice of parameters, the solution that is illustrated here is supercritical and exhibits bistability with the multiple roll oscillatory branch, as shown in Fig.~\ref{fig:convoscill}(b). This regular time-dependent solution takes the form of a standing wave oscillation in which the cells reverse their sign of vorticity every half period with approximately $16.2$ complete oscillations per diffusion time. The reversal of the vorticity continuously redistributes the magnetic field, regularly moving the flux sheets at the top and bottom of the layer into regions in which the flow converges.

Using the branch-tracking procedure, it is possible to locate a large number of different stable solutions in this system \cite{SB1999,SB1999b}. Fig.~\ref{fig:bifoscill} shows the stability boundaries for a selection of steady and oscillatory solution branches for these parameters, plotting the (average) Nusselt number, $\bar{N}$, on each branch as a function of $Q$. Note that we do not speculate on the location of the unstable solution branches \cite{SB1999}, since the location of these cannot be determined with any degree of accuracy using this particular method. This bifurcation diagram is by no means complete, with numerous other solution branches existing.  These have been omitted from this plot in order to highlight the branches of interest. In the context of present study, it is worth highlighting the presence of several solution branches corresponding to states which exhibit oscillatory behaviour. These are found at lower values of $Q$ than the spatially-periodic multiple roll oscillatory state that is illustrated in Fig.~\ref{fig:convoscill}(b). Indicated in the bifurcation diagram are 1, 2, 3 and 4-roll oscillatory solution branches as well as the 12 roll spatially-periodic state.

\begin{figure}
        \centering
        \includegraphics[scale=0.13]{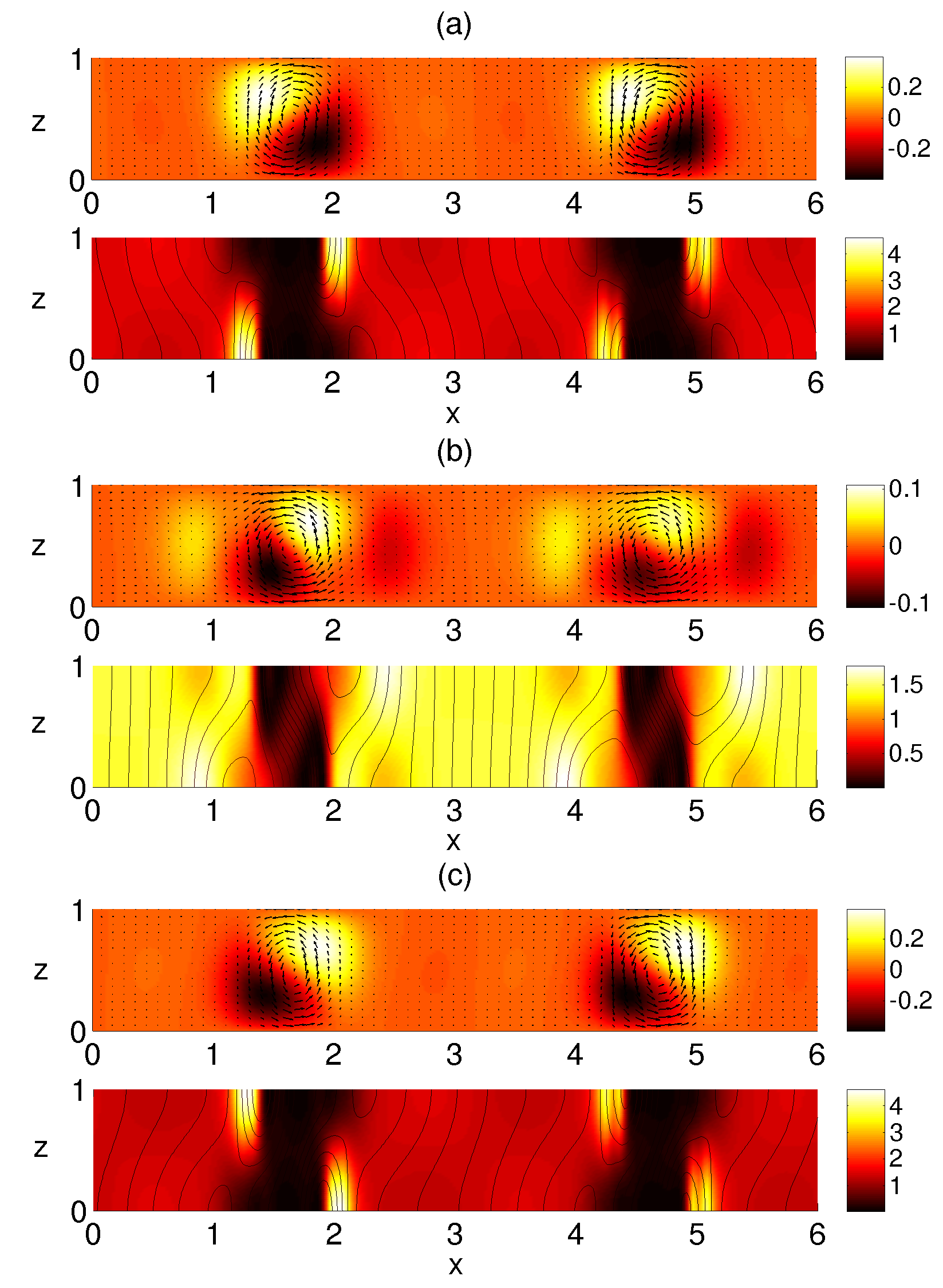}
        \caption{\label{fig:1D2O}(Color online) A time-sequence of states showing a half-period of oscillation (approximately $0.059$ diffusion times) for a two roll oscillatory solution at $R=20000$, $\sigma=1$, $\zeta=0.1$, $\lambda=6$ and $Q=18000$ ($Q_{\text{eff}} \approx 25480$).  The snapshots were taken at the peaks and troughs in the Nusselt number (a) $t=993.396$ (b) $t=993.425$ (c) $t=993.455$ (See Fig.~\ref{fig:1D2ONU}). Each of the three sets of plots shows the temperature perturbation $\theta$ (top) and magnetic field strength, $|{\boldsymbol B}|^2$ (bottom).}
        \vspace{0.33in}
        \includegraphics[scale=0.15]{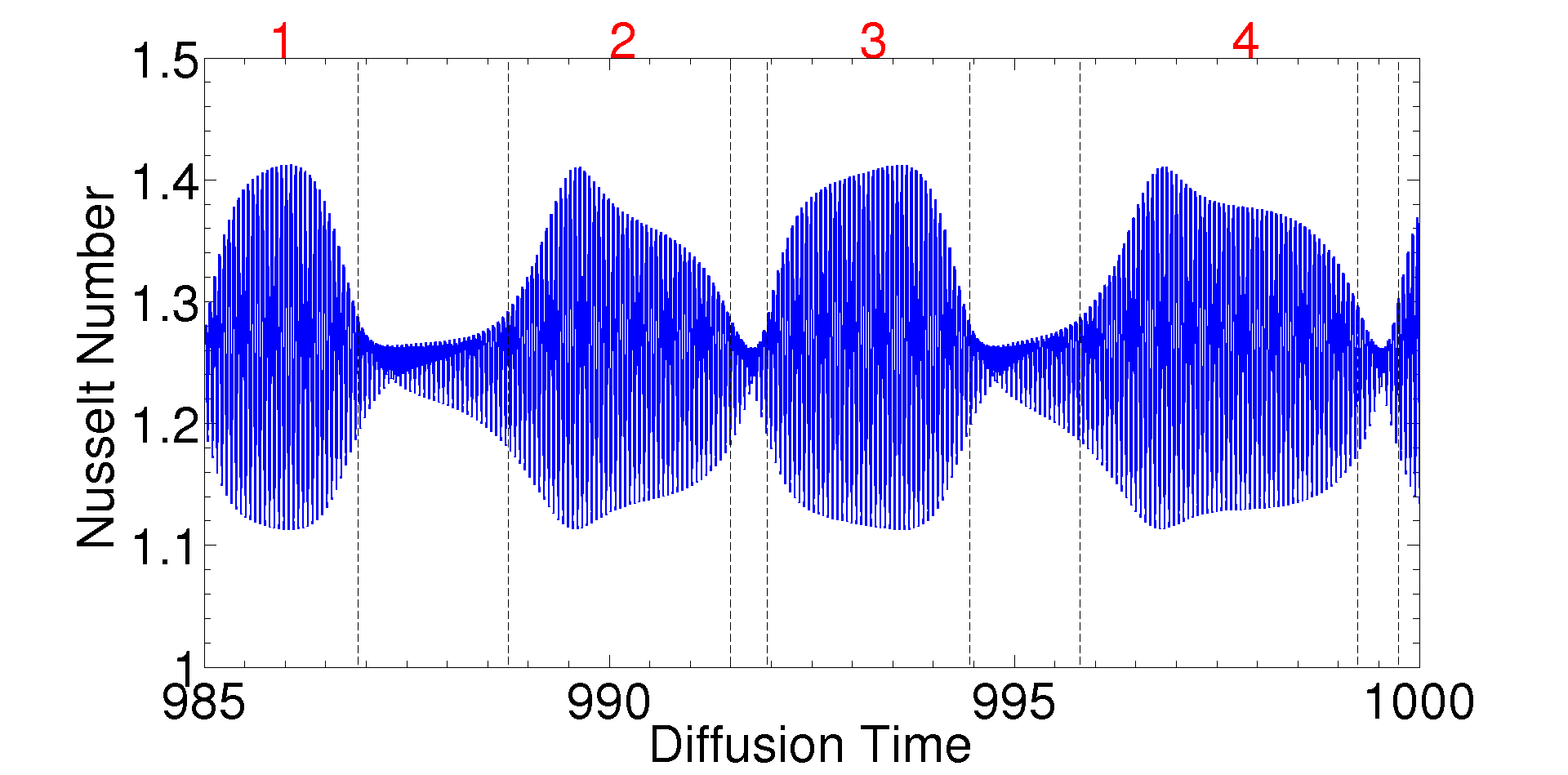}
        \caption{\label{fig:1D2ONU}(Color online) A time sequence of the Nusselt number for the two roll oscillatory state in Fig.~\ref{fig:1D2O}, indicating the modulated nature of the solution.  The average Nusselt number for this state is $\bar{N}=1.25$.}
      \end{figure}

Fig.~\ref{fig:1D2O} shows a time sequence of states (with time increasing downwards), illustrating one half-period of oscillation for a 2-roll oscillatory state at $Q=18000$. In this sequence, both rolls have the same sign of vorticity and oscillate in synchronisation. The sequence proceeds with both rolls decaying to a transitionary state in which the convective efficiency is lowest and the magnetic flux within the cells begins to redistribute.  The flux sheets at the top and bottom of the layer then switch to opposite sides of the cell and, as the convective efficiency begins to increase, they both develop into convective rolls with oppositely-signed vorticity to that with which they started.  This illustrates one half period of oscillation, corresponding to $8.6$ complete oscillations per diffusion time.  This solution, like all of the flux separated oscillatory states, is confined to a finite range of values for $Q$ ($14500 \le Q \le 19000$). 

      \begin{figure}
        \centering
        \includegraphics[scale=0.13]{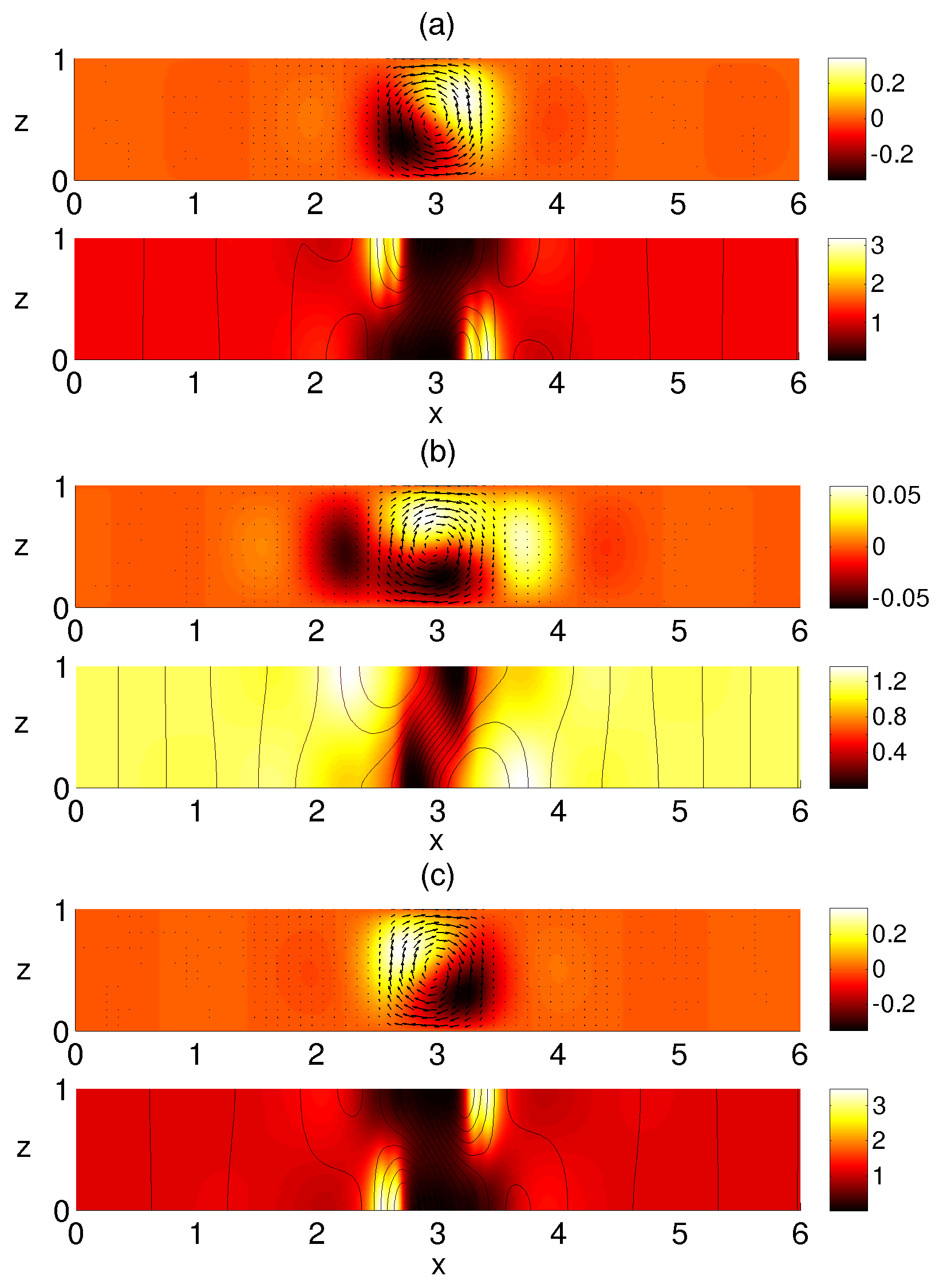}
        \caption{\label{fig:1D1O}(Color online) A time-sequence of states showing a half-period of oscillation (approximately $0.048$ diffusion times) for a single roll oscillatory solution at $R=20000$, $\sigma=1$, $\zeta=0.1$, $\lambda=6$ and $Q=22200$ ($Q_{\text{eff}} = 25247$).  The snapshots were taken at the peaks and troughs in the Nusselt number (a) $t=990.002$ ($N_{\text{max}}$) (b) $t=990.025$ ($N_{\text{min}}$) (c) $t=990.050$ ($N_{\text{max}}$). Each of the three sets of plots shows the temperature perturbation $\theta$ (top) and magnetic field strength, $|{\boldsymbol B}|^2$ (bottom).}%single roll oscillatory solution
        \vspace{0.2in}
        \includegraphics[scale=0.15]{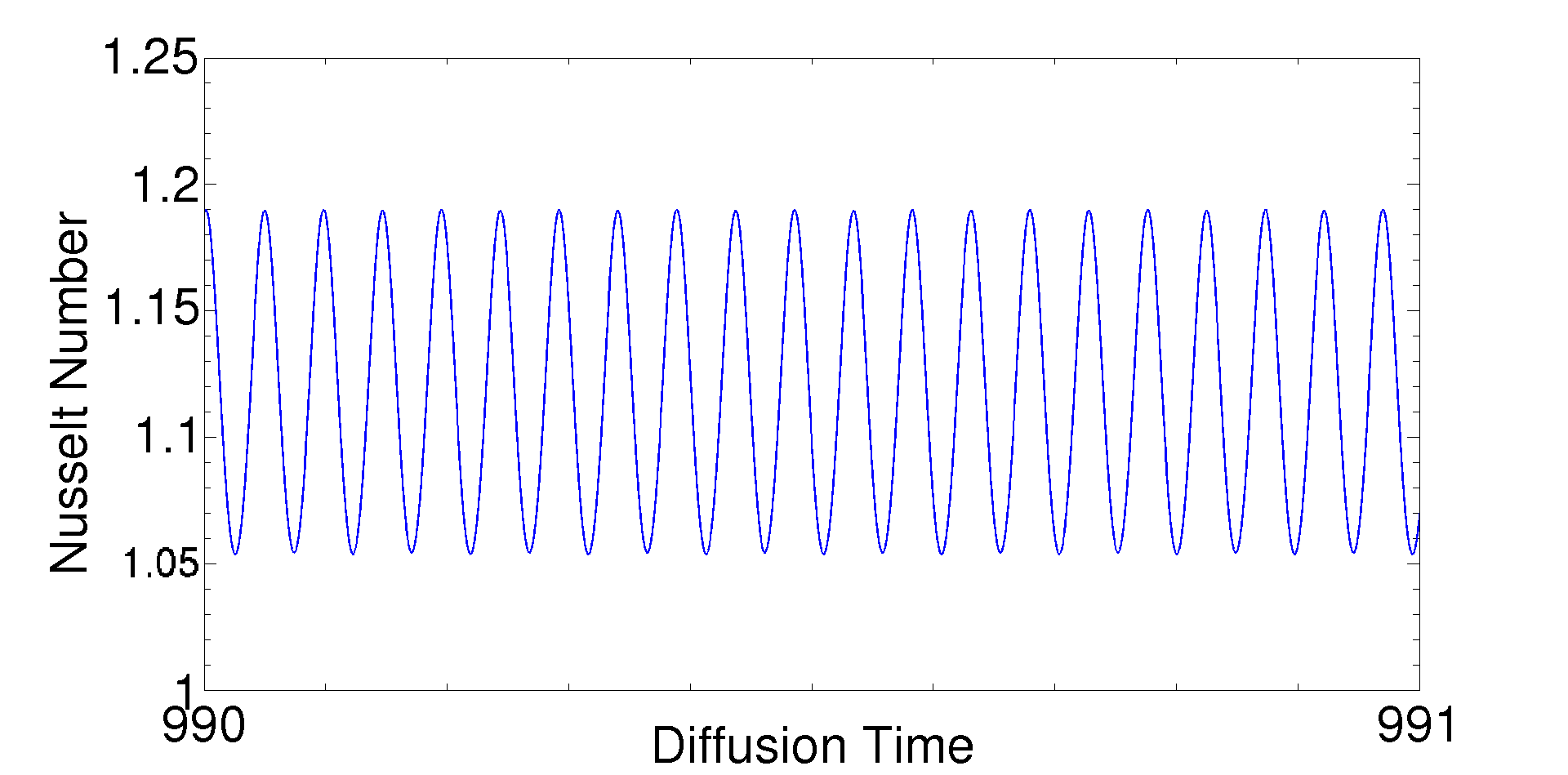}
        \caption{\label{fig:1D1ONU}(Color online) A time sequence of the Nusselt number for one full diffusion time for the localised oscillatory state in Fig.~\ref{fig:1D1O}.  There are approximately $10.4$ full oscillations per diffusion time with $N_{\text{max}}=1.19$, $N_{\text{min}}=1.05$ and $\bar{N}=1.12$.}
      \end{figure}

It is also worth noting that the multiple roll flux separated states, although they have a clear period of oscillation, are modulated over much longer time-scales.  Fig.~\ref{fig:1D2ONU} shows the variation of $N$, with time, for the state in Fig.~\ref{fig:1D2O}, over 15-diffusion times.  We observe that there are distinct phases, marked by the vertical dashed lines (- -), that this state transitions through, corresponding to different types of oscillation.  Phases one ($t \approx 985-986.9$) and three ($t \approx 992-994.5$) correspond to the state shown here with the cells oscillating in synchronisation, with the same sign of vorticity.  Phases two ($t \approx 988.8-991.5$) and four ($t \approx 995.8-999.2$) correspond to a state in which the two cells oscillate in synchronisation but with opposite vorticity to each other.  The transitionary regions between these states correspond to phases in which the oscillation of the cells is no longer synchronised.  The 3 and 4-roll states oscillate with similar modulated behaviour.

\begin{figure}
        \begin{center}
          \includegraphics[scale=0.11]{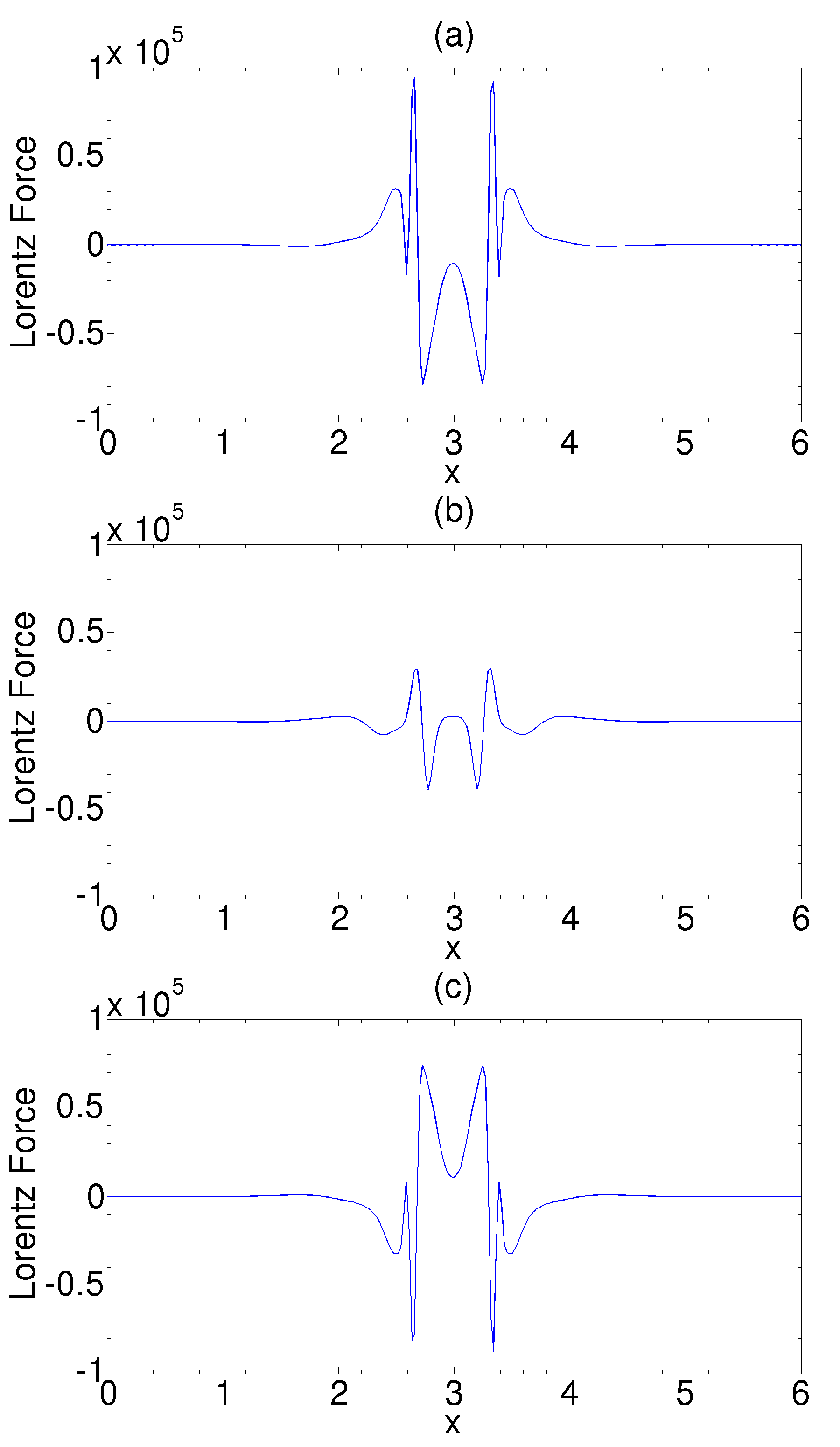}
          \caption{\label{fig:lor_for}(Color online) The corresponding Lorentz force plots for the snapshots in Fig.~\ref{fig:1D1O}.  The snapshots were taken at the peaks and troughs in the Nusselt number; (a) $t=990.002$ ($N_{\text{max}}$) (b) $t=990.025$ ($N_{\text{min}}$) (c) $t=990.050$ ($N_{\text{max}}$).}
        \end{center}
      \end{figure}

Taking the 2-roll oscillatory state at $Q = 18000$ and increasing the field strength, one can suppress one of the convective rolls so that only a single localised oscillatory cell remains. As indicated in Fig.~\ref{fig:bifoscill} the 2-roll solution branch loses stability at $Q\approx 19000$. By branch tracking, it can be shown that the 1o branch is stable in the range  $17900 \le Q \le 22200$, in close agreement with the findings of \cite{SB1999b} for this particular parameter regime. The localised oscillatory state at $Q=22200$ ($Q_{\text{eff}}=25247$) is illustrated in Fig.~\ref{fig:1D1O}. The cell oscillates in the same way as the 2-roll state, oscillating backwards and forwards, reversing the sign of the vorticity every half period (approximately $0.048$ diffusion times).  The oscillations in the Nusselt number are very regular (see Fig.~\ref{fig:1D1ONU}), with no modulation over large time-scales. Given that the Lorentz force plays a crucial role in driving the oscillations, it is instructive to look at the time-evolution of the spatial distribution of the Lorentz force. This is calculated as the sum of the terms proportional to $Q$ on the right hand side of Equation~(\ref{equ:w1cart}). Fig.~\ref{fig:lor_for} is a plot of the spatial-dependence of the Lorentz force terms for the snapshots shown in Fig.~\ref{fig:1D1O}. At each instant in time, the Lorentz force does appear to be nearly symmetric about $x=3$, although there are some small asymmetries in the distribution (visible primarily in the narrow spikes at the edge of the convecton in the upper and lower plots).  The symmetric nature of the Lorentz force distribution is probably due to the simplified nature of this model.  

We can see, from Fig.~\ref{fig:lor_for}, that the oscillatory state is not completely flux expelled, as the Lorentz force is non-zero within the cell.  This is in contrast to steady convectons, which are very efficient at expelling almost all magnetic flux from their interior. Within a cell that is completely devoid of magnetic flux, $A_{0}' = -1$, so that all terms that make up the Lorentz force in Equation~(\ref{equ:w1cart}) approximate to zero \cite{JD2007}. However, for the oscillatory convecton, indicated in Fig.~\ref{fig:1D1O}(a), at the peak of oscillation ($N_{\text{max}}$), when we expect a greater proportion of the flux to be expelled, we find that $A_{0}' \approx -0.67$, indicating that the cell is not completely field free. This is presumably due to the fact that convection is rather weak ($N_{\text{max}}=1.19$). The inefficiency of flux expulsion is almost certainly essential for the persistence of oscillations in these localised states, and also presumably explains their susceptibility to convective perturbations to the outside region of the cell, leading to their narrow range of stability. More specifically at $Q \approx 22300$ the state described above becomes unstable, and the solution transitions to the 12-roll oscillatory solution indicated in Fig.~\ref{fig:convoscill}(b). As $Q$ is decreased along the 1o branch, although the motion is still dominated by a single oscillatory roll, weak convective motions start to appear within the magnetically-dominated region of the domain. This occurs at approximately $Q \approx 20500$ (at which point $Q_{\text{eff}} \approx Q_{\text{max}}^{\text{(o)}}$). As $Q$ is decreased further the state becomes increasingly perturbed and the state is eventually overcome by these oscillations at $Q \approx 17800$ ($Q_{\text{eff}}=19852$), transitioning to the 2-roll state indicated in Fig.~\ref{fig:1D2O}. Fig.~\ref{fig:1d1oscill_NL} shows the solution right at the lower end of the 1o solution branch. 

      \subsubsection{\label{ssec:PDOOS}Varying the system parameters}

In this section we assess the effects of varying the Rayleigh number, $R$, the diffusivity ratio, $\zeta$, and the aspect ratio, $\lambda$. The aim here is to determine whether or not these localised oscillatory states are restricted to the particular parameter regime that was discussed above. This is the first time that the parameter-dependence of these states has been explored. In fact, these solutions do appear to be reasonably robust, in the sense that similar solutions can be found across a range of different parameters. In Table~\ref{tbl:summary} we summarise some of the findings of this parametric survey. Various properties of the oscillatory state are given for $Q_{\text{max}}$ (the upper end of the solution branch): $\bar{\lambda}$ corresponds to the mean width of the cell, whilst $N_{\text{max}}$ and $\bar{N}$ give the peak and mean values of the Nusselt number (respectively). Also indicated is the location of the linear stability boundary for the trivial state, $Q_{\text{max}}^{\text{(o)}}$ \cite{MPNW1982}.

      \begin{figure}
        \begin{center}
          \includegraphics[scale=0.14]{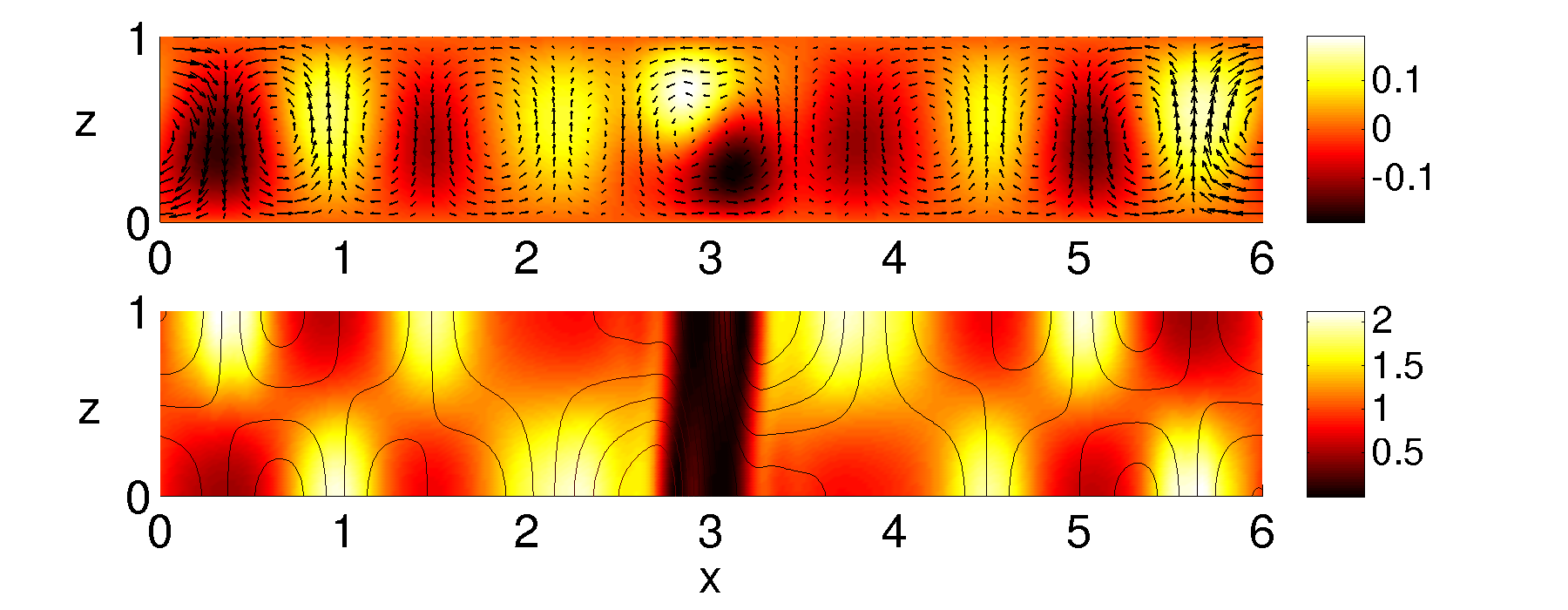}
          \caption{\label{fig:1d1oscill_NL}(Color online) A snapshot of a solution on the single roll oscillatory solution branch at $R=20000$, $\sigma=1$, $\zeta=0.1$, $\lambda=6$ and $Q=17900$, clearly indicating the perturbations to the outside region of the cell that develop as $Q$ is decreased.  This state was taken at an irregular spike in the Nusselt number ($N_{\text{max}}=1.34$) in order to highlight the eddies in the outside region at the peak in their intensity ($\bar{N}=1.20$).  The plot shows the temperature perturbation $\theta$ (top) and magnetic field strength, $|{\boldsymbol B}|^2$ (bottom).}
        \end{center}
      \end{figure}

      \begin{figure}
        \begin{center}
        \includegraphics[scale=0.14]{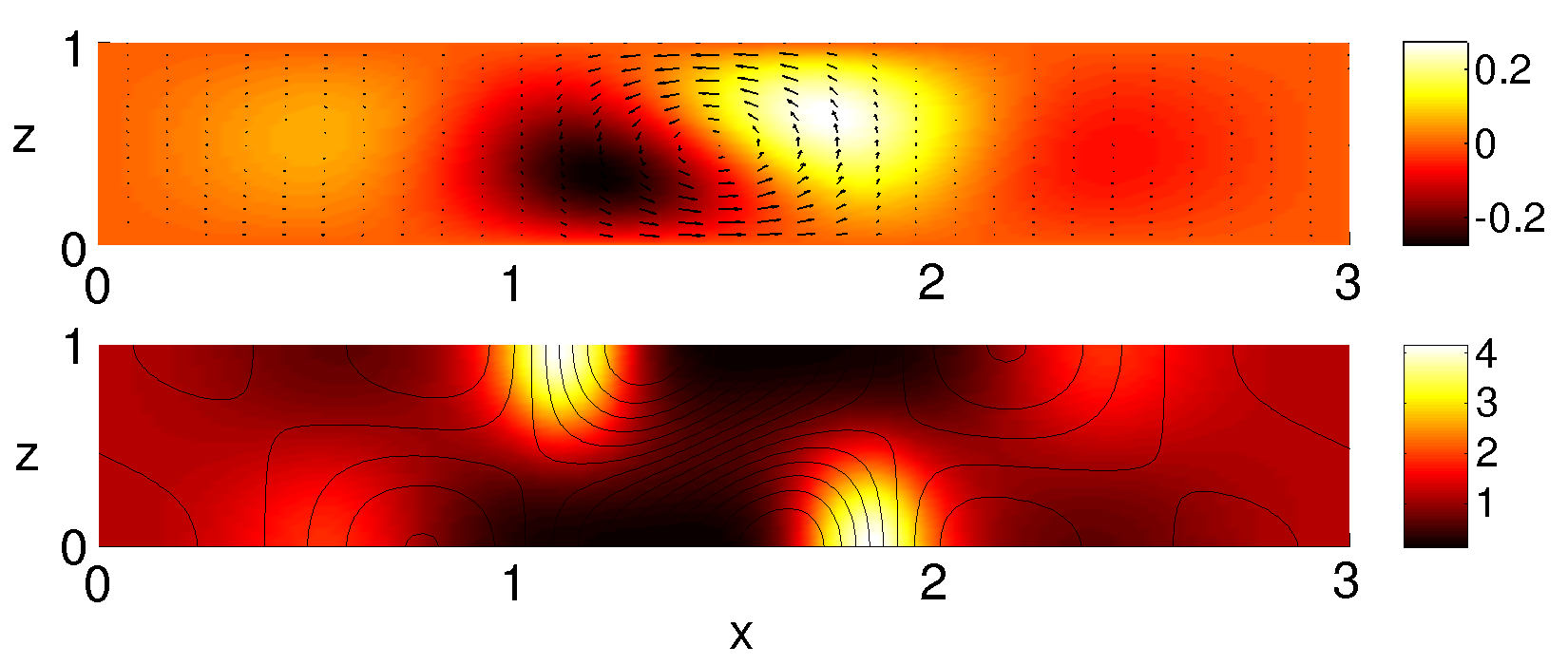}
        \caption{\label{fig:lambda3}(Color online) A snapshot of a solution on the single roll oscillatory solution branch at $R=20000$, $\sigma=1$, $\zeta=0.1$, $\lambda=3$ and $Q=19800$ ($Q_{\text{eff}}=22910$), indicating that in a restricted domain the structure of the localised oscillatory cell remains consistent with that in a wider domain. This state was taken at the peak in the Nusselt number ($N_{\text{max}}=1.21$) where $\bar{N}=1.13$ for this state.  The plot shows the temperature perturbation $\theta$ (top) and magnetic field strength, $|{\boldsymbol B}|^2$ (bottom).}
        \end{center}
      \end{figure}

      \begin{table}[ht]
        \caption{\label{tbl:summary} A summary of the stability range of oscillatory convectons for a range of values of $\lambda$, $\zeta$ and $R$.  Indicated are a number of properties of the cells at $Q_{\text{max}}$: $\bar{\lambda}$ corresponds to the mean width of the cell, whilst $N_{\text{max}}$ and $\bar{N}$ give the peak and mean values of the Nusselt number (respectively).  Also indicated is the location of the largest oscillatory Hopf bifurcation $Q_{\text{max}}^{\text{(o)}}$. Note that the period of oscillation is given in terms of diffusion times.}%(in brackets we give the number of oscillations per diffusion time)  
        \centering
        \begin{tabular}{c c c c c c c c c c}
          \hline\hline \\[-2.0ex] 
          $\lambda$ & $\zeta$ & $R$ & $Q_{\text{min}}$ & $Q_{\text{max}}$ & $Q_{\text{max}}^{\text{(o)}}$ & $\bar{\lambda}$ & $N_{\text{max}}$ & $\bar{N}$ & Period \\ [0.5ex] 
          \hline \\[-2.0ex]
          6 & 0.1 & 5000 & 3400 & 4200 & 4220 & 0.803 & 1.14 & 1.08 & 0.248 \\%
          6 & 0.1 & 10000 & 7900 & 10300 & 10447 & 0.664 & 1.18 & 1.10 & 0.161 \\%
          6 & 0.1 & 20000 & 17900 & 22200 & 24047 & 0.599 & 1.19 & 1.12 & 0.097 \\%
          6 & 0.1 & 50000 & 47600 & 51000 & 68800 & 0.531 & 1.40 & 1.26 & 0.057 \\%
          6 & 0.2 & 20000 & 7900 & 9800 & 10727 & 0.592 & 1.19 & 1.12 & 0.116 \\%
          6 & 0.3 & 20000 & 4700 & 5800 & 6420 & 0.587 & 1.19 & 1.11 & 0.120 \\%
          3 & 0.1 & 20000 & 14400 & 19800 & 24047 & 0.491 & 1.21 & 1.13 & 0.068 \\%
          2 & 0.1 & 20000 & 11700 & 18100 & 24047 & 0.441 & 1.27 & 1.18 & 0.068 \\%
          1.5 & 0.1 & 20000 & 8800 & 17100 & 21905 & 0.464 & 1.44 & 1.26 & 0.067 \\%
          8 & 0.1 & 20000 & 18800 & 23300 & 24191 & 0.629 & 1.15 & 1.09 & 0.108 \\%
          16 & 0.1 & 20000 & 19900 & 24256 & 24257 & 0.610 & 1.08 & 1.05 & 0.122 \\%
          \hline\hline
        \end{tabular}
      \end{table}

One of the key findings of this survey is that the existence of the oscillatory convecton is found to be critically dependent upon $\zeta$.  Fixing $R=20000$, $\sigma=1$ and $\lambda=6$ we find that these states can be found only for $\zeta < 0.6$. It is found that as $\zeta$ is increased, there is a reduction in the stability range of these states, as indicated in Table \ref{tbl:summary}.  Steady convectons exhibit a similar $\zeta$ dependence \cite{JD2007}. Where oscillatory localised states do exist, the cell width appears to be relatively insensitive to the precise choice of $\zeta$, although we see a marked increase in the period of oscillation at higher values of $\zeta$. This is presumably due to the fact that the lower $\zeta$ oscillatory convectons are stable at higher values of $Q$, which increases the amplitude of the Lorentz force in the boundary layers.

For steady localised states the effect of increasing the box width is to shift the stability range to higher values of $Q$ \cite{SB1999b}.  We observe the same phenomenon here for the oscillatory convectons, although the actual branch width is always relatively small.  For steady convectons this branch moves completely into the subcritical regime for $\lambda \gg 1$, with the lower end of the steady convecton branch positioned at $Q_{\text{max}}^{\text{(o)}}$ \cite{SB1999}.  The reason for this is that the cell is so small in the wide domain that the increase in flux in the outside region is not large enough to suppress all convective perturbations if $Q<Q_{\text{max}}^{\text{(o)}}$. In this oscillatory case, the upper limit of the stability range (in $Q$) of the localised state seems to be tending towards $Q_{\text{max}}^{\text{(o)}}$ as the box size increases, with almost no variation in the branch width, as can be seen from Table \ref{tbl:summary}. Furthermore, we find no evidence for the existence of subcritical oscillatory convectons. This is specifically highlighted by the $\lambda=16$ case in Table \ref{tbl:summary}, which indicates that the solution decays rapidly to the trivial state almost exactly at $Q_{\text{max}}^{\text{(o)}}$. Unlike steady convectons, oscillatory convectons can also be found in much smaller domains. As indicated in Table \ref{tbl:summary}, the localised oscillatory state for $\lambda=3$ (which is illustrated in Fig.~\ref{fig:lambda3}) is stable in the range $14400 \le Q \le 19800$. At $Q=18000$, $\bar{N}=1.25$ and $N_{\text{max}}=1.41$, both of which compare very favourably to the $\lambda=6$ two roll state that is shown in Fig.~\ref{fig:1D2O}. Similarly we may compare the Nusselt number for a single roll oscillatory state in a domain of size $\lambda=6/3=2$ and $\lambda=6/4=3/2$ which oscillate about average Nusselt numbers of $\bar{N}=1.36$ and $\bar{N}=1.51$ respectively, in agreement with those for the three ($\bar{N}=1.37$) and four roll ($\bar{N}=1.50$) oscillatory states in the $\lambda=6$ domain. We conclude from this that the 2o, 3o and 4o states in the $\lambda=6$ domain could be regarded simply as collections of interacting oscillatory convectons (any of which could exist in isolation). 

Varying the Rayleigh number at fixed $\zeta$ (which is equivalent to varying the strength of the convective driving forces) we find localised oscillatory convectons for $5000 \le R \le 50000$, however, for larger values of $R$ (e.g. $R=70000$), stable states could not be found.  For $R=50000$, oscillatory convectons are stable in the range $47600 \le Q \le 51000$, however, even at the upper boundary ($Q_{\text{max}}$) of the solution branch ($Q \approx 51000$, $N_{\text{max}}=1.42$, $\bar{N}=1.28$, $Q_{\text{max}}^{\text{(o)}}=68800$), small counter rotating eddies either side of the main cell are present, indicated by the fact that $Q_{\text{eff}} \approx 56848 < Q_{\text{max}}^{\text{(o)}}$, similar to the state in Fig.~\ref{fig:1d1oscill_NL}.  These perturbations outside of the cell for higher Rayleigh numbers are most likely the result of the inability of these states to expel significant quantities of magnetic flux. As a result, significant increases in the thermal forcing mean that the field is not locally strong enough to suppress all convective perturbations and stabilise the layer. This could give a possible explanation as to why these states have not been observed at higher Rayleigh numbers.

    \subsection{\label{ssec:OSTFRM}The Fully Resolved Model}

      \begin{figure}
        \begin{center}
        \includegraphics[scale=0.14]{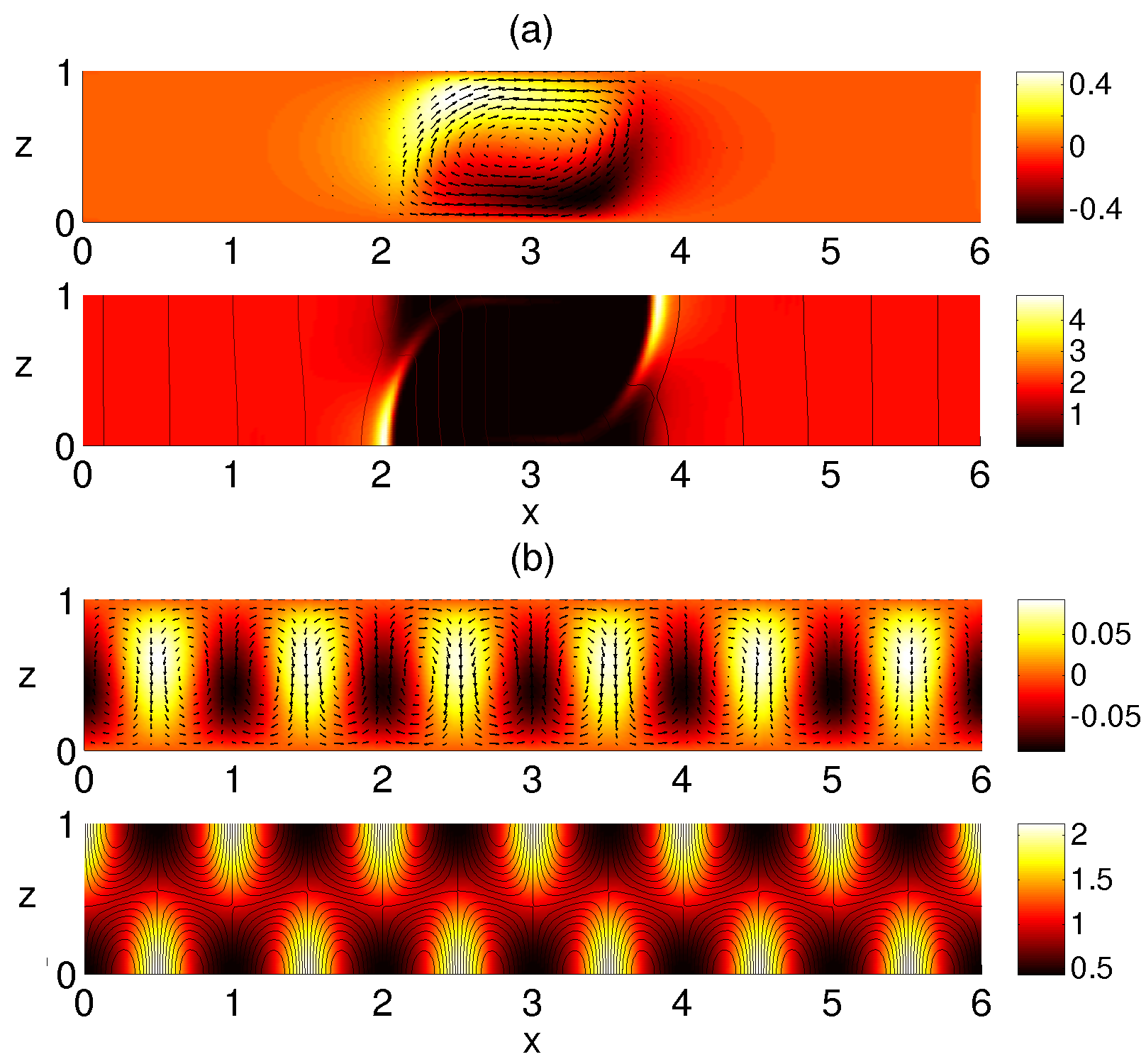}
        \caption{\label{fig:convoscill2d}(Color online) Two supercritical solutions obtained for the same parameter values, $R=20000$, $\zeta=0.1$, $\sigma=1$, $\lambda=6$ with $Q=22000$, for the fully-resolved model, indicating the bistability of these states; (a) a convecton solution with $N=2.38$ and $\bar{\lambda}=1.94$ ($Q_{\text{eff}}=48048$) (b) a 12-roll oscillatory state with $\bar{N}=1.09$. Each plot shows the temperature perturbation $\theta$ (top) and magnetic field strength, $|{\boldsymbol B}|^2$ (bottom).}
        \end{center}
      \end{figure}

Having analysed the truncated model in some detail, we now turn to the model of fully-resolved two-dimensional magnetoconvection. In any model of this type, the presence of slowly decaying transients implies that long time integrations are required in order to ensure the persistence of localised states, so that the addition of the extra spatial structure makes it impractical to carry out detailed parametric surveys of the fully-resolved system. Therefore our parameter choices must be guided by the results from the truncated model. However, it is not always the case that all behaviour observed in a simplified model translates to the fully-resolved case. For example, consider the $R=20000$, $\zeta=0.1$, $\sigma=1$ and $\lambda=6$ case. In the case of steady convectons, simulations show a very dramatic increase in the stability range, from $13700 \le Q \le 24300$ in the truncated model to $6000 \le Q \le 25800$ in the fully-resolved model.  Another trait of the fully-resolved simulations is that for a particular set of the parameters, the cell size and convective efficiency is significantly increased in comparison to the truncated model.  The increased cell size can be seen in Fig.~\ref{fig:convoscill2d}, which shows a fully-resolved steady convecton, found by branch tracking, as well as a 12-roll oscillatory state, for $Q=22000$.  This should be compared with Fig.~\ref{fig:convoscill}, which shows the results from the truncated model for the same set of parameters. The Nusselt number in this case is $N=2.38$ (as opposed to $N=1.31$), an increase which can mostly be attributed to the increased cell size. Nonetheless, despite the apparent differences in the solutions, the presence of a steady convecton for these parameter values suggests that it is not unreasonable to suppose that the truncated model should provide a reasonable indication of the appropriate regions of parameter space to search for these solutions in this fully-resolved model. 

      \begin{figure}
        \begin{center}
          \includegraphics[scale=0.14]{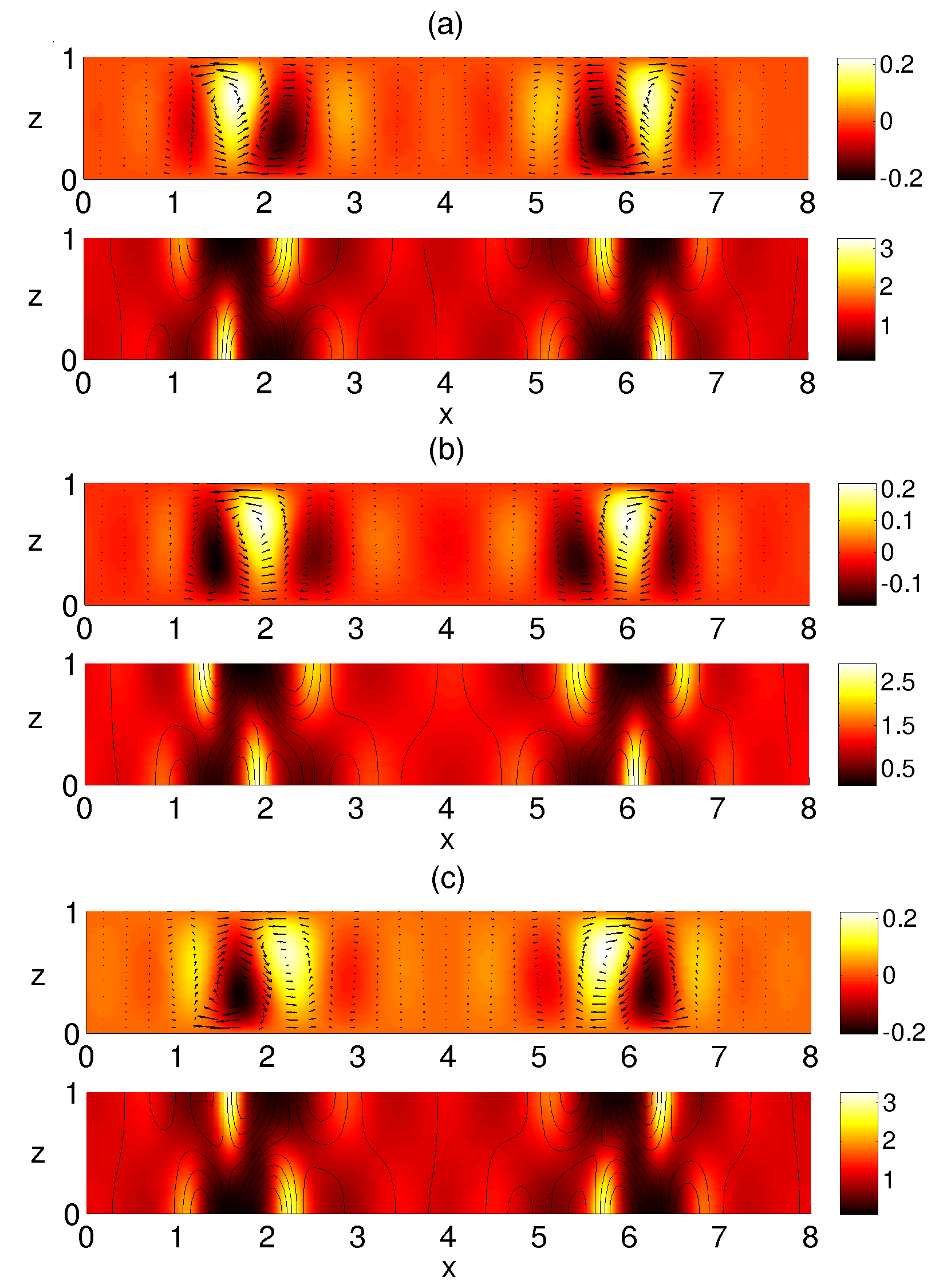}
          \caption{\label{fig:2D2O}(Color online) A time sequence of convective states for a two-roll flux separated oscillatory state, showing one half period of oscillation (approximately $0.034$ diffusion times) at $R=20000$, $\sigma=1$, $\zeta=0.1$, $\lambda=8$ and $Q=21000$. The snapshots were taken at equally spaced time-intervals; (a) $t=1543.005$ (b) $t=1543.022$ (c) $t=1543.039$. Each of the three plots shows the temperature perturbation, $\theta$, (top) and magnetic field strength, $|{\boldsymbol B}|^2$, (bottom).}
        \end{center}
      \end{figure}

      \begin{figure}
        \centering
        \includegraphics[scale=0.14]{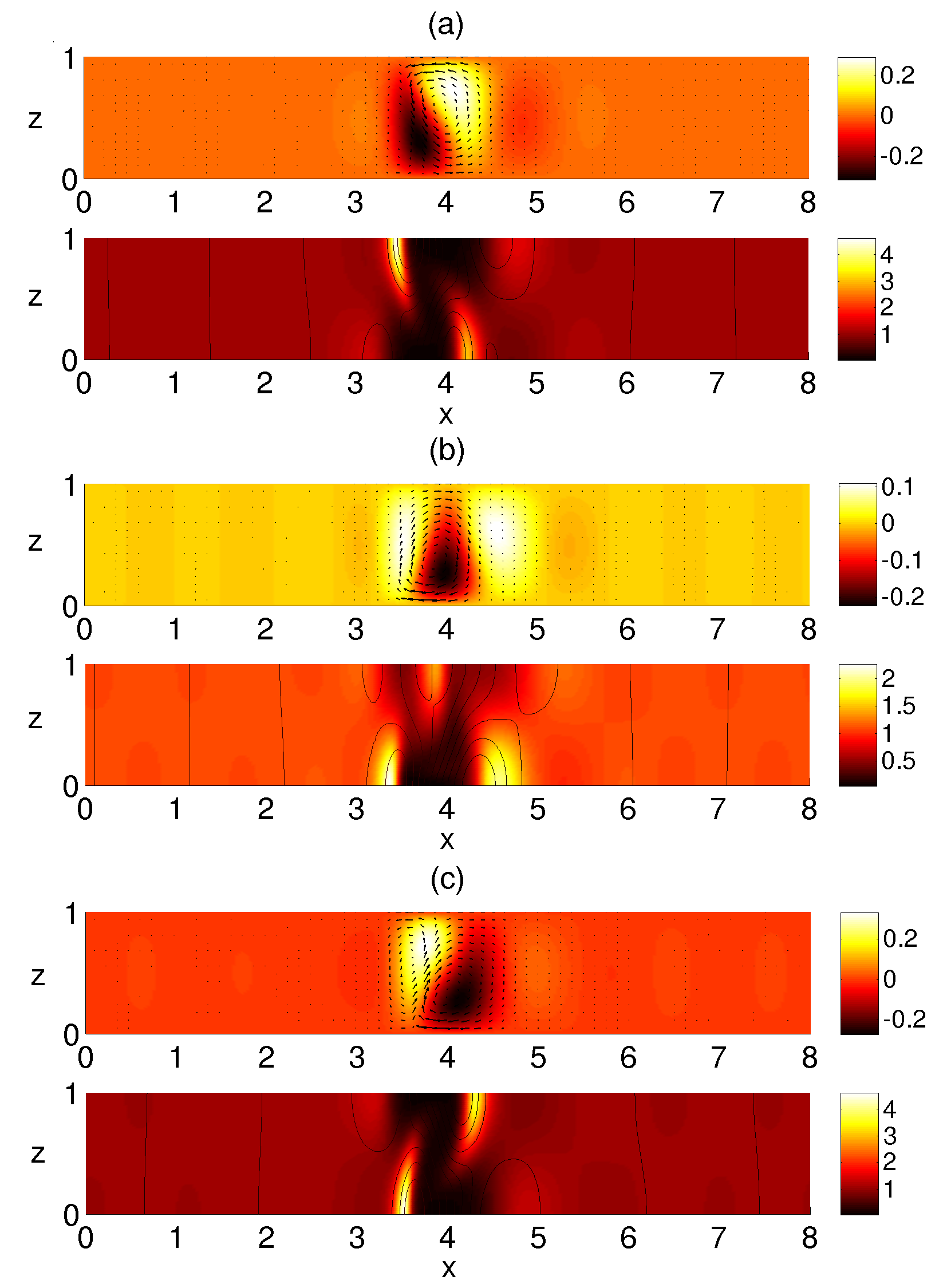}
        \caption{\label{fig:2D1O}(Color online) A time sequence of states for one reversal in the vorticity of the state (approximately $0.046$ diffusion times) at $R=20000$, $\sigma=1$, $\zeta=0.1$ and $Q=22000$.  The width of the cell at the peak of oscillation is $\bar{\lambda}=0.787$.  The snapshots were taken at the peaks and troughs in the Nusselt number; (a) $t=1499.001$ ($N_{\text{max}}$) (b) $t=1499.027$ ($N_{\text{min}}$) (c) $t=1499.047$. Each of the three sets of plots shows the temperature perturbation $\theta$ (top) and magnetic field strength, $|{\boldsymbol B}|^2$ (bottom).}
        \vspace{0.2in}
        \includegraphics[scale=0.15]{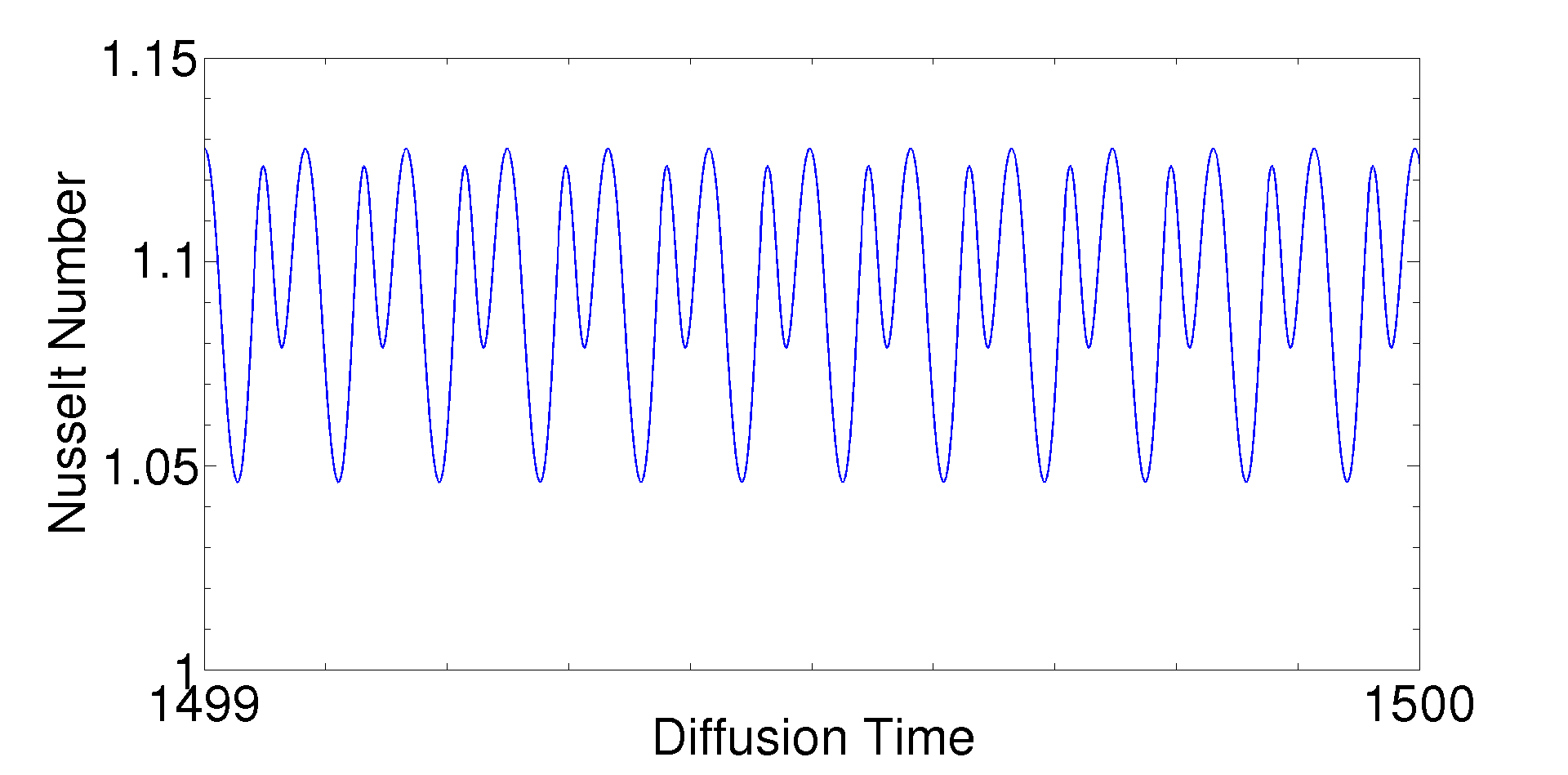}
        \caption{\label{fig:2D1ONu}(Color online) Time-evolution of the Nusselt number for the localised oscillatory cell shown in Fig.~\ref{fig:2D1O}, over approximately one full diffusion time.  Note that there are approximately $12.049$ full oscillations per diffusion time with $N_{\text{max}}=1.13$, $N_{\text{min}}=1.05$ and $\bar{N}=1.09$.}
      \end{figure}

Initial attempts were made to locate oscillatory convectons for $R=20000$, $\sigma=1$, $\zeta=0.1$ and $\lambda=6$. However, branch-tracking by increasing $Q$ through the flux-separated oscillatory states failed to produce the required solution. Tracking down from higher $Q$ (in the other direction), we found that the spatially-periodic oscillatory state transitions directly to a 2-roll oscillatory state. So we were unable to find a stable oscillatory convecton by following this branch-tracking process. However, it was noticed that there was a transient localised cell during the transition to the 2-roll state. Although clearly a transient, this localised state was present for approximately 50 diffusion times, which highlights the need for long time integrations in systems of this type. Given the presence of this long-lived transient, it is certainly possible that a stable oscillatory convecton solution branch does exist, but perhaps over a narrower region of parameter space than the corresponding state in the truncated model. 

Given that steady convectons tend to be wider in this fully-resolved model, it is not unreasonable to assume that the stability of these solution branches might depend crucially upon the aspect ratio. Increasing the box width to $\lambda=8$, we found a two roll oscillatory state at $Q=21000$ (which is illustrated in Fig.~\ref{fig:2D2O}). In this fully-resolved case, the oscillations no longer take the form of localised standing waves. Instead, the oscillation has more of a horizontal component. This can be seen from the structure of the magnetic boundary layers which move gradually across the cell. The translational velocities are to the right in the left hand cell and to the left in the right hand cell.  As was the case in the truncated model, the Nusselt number variation of this 2-roll state is modulated over a large time-scale, and there are distinct phases (of finite duration) in which the cells oscillate in a synchronous way. 

\begin{figure}
        \begin{center}
          \includegraphics[scale=0.11]{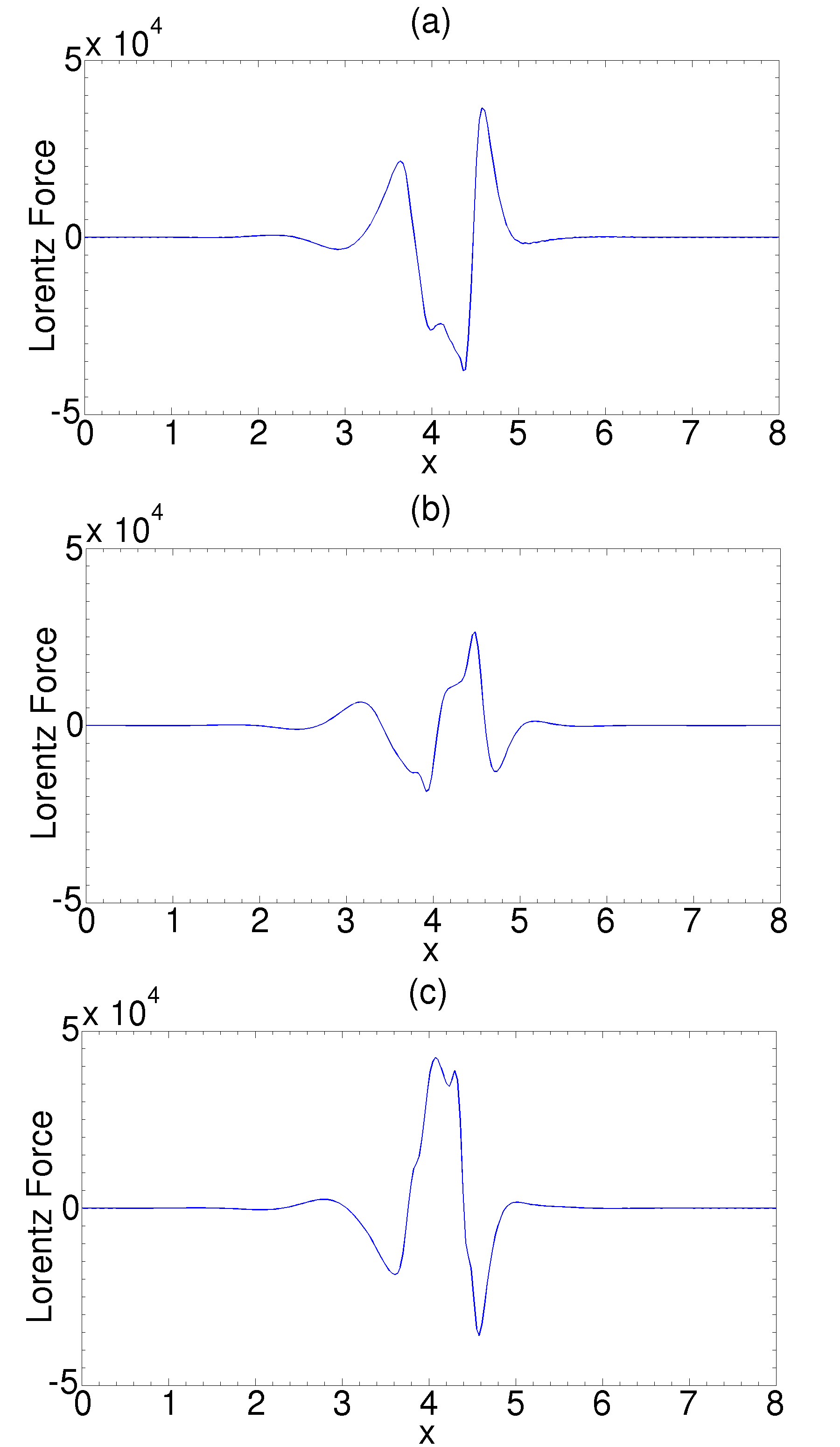}
          \caption{\label{fig:2D1Olor}(Color online) The corresponding Lorentz force plots for the snapshots in Fig.~\ref{fig:2D1O}.  The snapshots were taken at consecutive peaks and troughs in the Nusselt number, at the mid-layer of the box ($z=0.5$); (a) $t=1499.001$ ($N_{\text{max}}$) (b) $t=1499.027$ ($N_{\text{min}}$) (c) $t=1499.047$.}
        \end{center}
      \end{figure}

Taking the 2-roll state, we increased the field strength and found that a single roll oscillatory localised state stabilised at $Q=22000$ ($Q_{\text{eff}}=24127$), as shown in Fig.~\ref{fig:2D1O}.  To verify that this state was not a transient phenomenon, it was evolved for over $1000$ diffusion times, carefully monitoring the mean Nusselt number, which maintains a stable value of $\bar{N}=1.09$. This average Nusselt number is in fact slightly lower than that for the comparable state in the truncated model. The actual time-evolution of the Nusselt number is less trivial than in the corresponding solution in the truncated model. As shown in Fig.~\ref{fig:2D1ONu}, whilst still periodic, it is clear that the Nusselt number has a more complex functional form. Furthermore, the maximum Nusselt number ($N_{\text{max}}=1.13$) is comparatively large. This behaviour is a consequence of the fact that the mode of oscillation in the fully-resolved case does not take the form of a localised standing wave, which implies that there is an asymmetry between the vertical distribution of the hot and cold regions of the oscillatory convective motion. The Lorentz force, which is calculated as the magnitude of the third term on the right hand side of Equation~(\ref{equ:NS}), is indicated in Fig.~\ref{fig:2D1Olor} for this state. The Lorentz force is calculated at the mid-layer of the box ($z=0.5$), where the vorticity takes its maximum value, in agreement with the dominant modes of the truncation, so that the Lorentz force can be directly compared between the two models. We observe an asymmetry in the profile here which is in contrast to the very symmetric form in the truncated model and is most likely responsible for the differences between their modes of oscillation.  Due to the numerical complexity of these simulations we have not carried out a detailed parametric survey for these fully-resolved localised states. However, we have found that these states exist in the (approximate) range $19000 \le Q \le 22000$.

  \section{\label{sec:Conclusions}Conclusions}

We have demonstrated the existence of oscillatory localised states in fully-resolved two-dimensional magnetoconvection. This is the first time that a localised oscillatory state of this kind has been found in a fully-resolved magnetoconvection simulation. The time-dependence of these states is found to be more complex than that observed in simplified models, differing from a localised standing wave due to the presence of asymmetries in the Lorentz force. These states were located due to an understanding of their stability properties that we gained by performing a parametric survey using a vertically truncated model. These oscillatory solutions are found to be restricted to the low $\zeta$ regime, existing only for $\zeta < 0.6$. However, these states are found to be much more robust to variations in the Rayleigh number, with stable states certainly existing in the range $5000 \le R \le 50000$, although there is no evidence for their existence at larger values. These oscillatory localised states expel magnetic flux less efficiently than their steady counterparts, so they are never entirely field free. This probably explains why they do not seem to exist in the subcritical regime. This point was highlighted by increasing the aspect ratio of the domain, a process which moved the upper stability boundary of these states closer to $Q_{\text{max}}^{\text{(o)}}$ without a corresponding increase in the branch width. For a box with aspect ratio $\lambda=16$ it was found that the oscillatory convecton branch terminated almost precisely at $Q_{\text{max}}^{\text{(o)}}$, suggesting that these solutions can never exhibit subcritical behaviour. Thus we hypothesise that the existence of such states is a consequence of the finite geometry of the box. Having said that, reducing the box size appears to have no adverse effect upon the existence of these oscillatory localised states. 

Of course, one of the main limitations of this study is that it is restricted to two spatial dimensions, and the localised solutions that have been found would correspond to roll-like structures in a three-dimensional system. These isolated rolls would almost certainly be unstable to three-dimensional convective perturbations, so any localised oscillatory states in three spatial dimensions would probably have a more complicated structure than these simple rolls. Time-dependent plume-like structures have been found in vertically-truncated three-dimensional Boussinesq magnetoconvection \cite{SBNW2002} so, given the findings of this paper, it is not unreasonable to assume that fully three-dimensional Boussinesq magnetoconvection should exhibit similar behaviour. Further work is needed to confirm whether or not this is the case. It has already been established that steady convectons can be found in three-dimensional compressible magnetoconvection \cite{SHPB2011}. Work is in progress to determine the effects of stratification upon both steady and oscillatory localised states in two and three spatial dimensions. 

\bibliography{references} 

\end{document}